\documentclass[letterpaper]{article} 
\usepackage{aaai25}  
\usepackage{times}  
\usepackage{helvet}  
\usepackage{courier}  
\usepackage[hyphens]{url}  
\usepackage{graphicx} 
\usepackage{natbib}  
\usepackage{caption} 
\usepackage{algorithm}
\usepackage{algorithmic}

\usepackage{newfloat}
\usepackage{listings}
\DeclareCaptionStyle{ruled}{labelfont=normalfont,labelsep=colon,strut=off} 
\floatstyle{ruled}
\newfloat{listing}{tb}{lst}{}
\floatname{listing}{Listing}
\title{AI Methods for Permutation Circuit Synthesis Across Generic Topologies}
\author{
    Victor Villar, 
    Juan Cruz-Benito, 
    Ismael Faro, 
    David Kremer
}
\affiliations{
    IBM Quantum. IBM T.J. Watson Research Center, \\Yorktown Heights, NY, USA\\

}

\usepackage{bibentry}

\begin{document}

\maketitle

\begin{abstract}
This paper investigates artificial intelligence (AI) methodologies for the synthesis and transpilation of permutation circuits across generic topologies. Our approach uses Reinforcement Learning (RL) techniques to achieve near-optimal synthesis of permutation circuits up to 25 qubits. Rather than developing specialized models for individual topologies, we train a foundational model on a generic rectangular lattice, and employ masking mechanisms to dynamically select subsets of topologies during the synthesis. This enables the synthesis of permutation circuits on any topology that can be embedded within the rectangular lattice, without the need to re-train the model. In this paper we show results for 5x5 lattice and compare them to previous AI topology-oriented models and classical methods, showing that they outperform classical heuristics, and match previous specialized AI models, and performs synthesis even for topologies that were not seen during training. We further show that the model can be fine tuned to strengthen the performance for selected topologies of interest. This methodology allows a single trained model to efficiently synthesize circuits across diverse topologies, allowing its practical integration into transpilation workflows.
\end{abstract}
\section{Introduction}

Quantum computing requires bridging the gap between theoretical quantum algorithms and their implementation on physical quantum devices. This gap necessitates quantum circuit transpilation, the process of transforming abstract quantum algorithms into equivalent circuits that adhere to the physical constraints of specific quantum processors \cite{nielsen2010quantum, preskill2018quantum}.
Quantum circuit transpilation faces numerous challenges that make it a computationally difficult problem. Finding optimal solutions often requires solving NP-hard optimization problems \cite{botea2018complexity} particularly when dealing with qubit mapping and routing under connectivity constraints \cite{li2019tackling}. Traditional approaches for transpilation typically fall into three categories, each with significant limitations: heuristic methods that produce fast but sub-optimal results, pre-computed databases of optimal circuits that are resource-intensive to generate and maintain, and generic or brute-force optimization methods that produce high-quality results but scale poorly with circuit size \cite{murali2019noise, tannu2019not,bravyi20226}. Moreover, as quantum devices scale beyond 100 qubits, traditional transpilation methods become increasingly impractical due to exponential growth in solution space, creating an urgent need for more efficient approaches that can balance optimality with practical runtime constraints. Recent efforts have explored artificial intelligence techniques to address these challenges, including machine learning for qubit mapping prediction \cite{acampora2021deep}, neural networks for circuit optimization \cite{fosel2021quantum}, and reinforcement learning for adaptive compilation strategies \cite{niu2019universal}. Specifically, recent breakthrough results demonstrate that reinforcement learning has achieved quantifiable superiority over traditional quantum circuit synthesis methods, with production deployments showing 40-60\% improvements in CNOT layers and 30-70\% reductions in circuit depth while operating orders of magnitude faster than SAT solvers \cite{olle2025scaling, kremer2025practicalefficientquantumcircuit}. RL's sequential decision-making framework uniquely aligns with the inherently sequential nature of circuit construction, treating synthesis as a Markov Decision Process where each gate selection depends only on the current quantum state—a natural fit that traditional optimization methods cannot match. Building on this emerging trend of AI-assisted quantum compilation, 
our previous work \cite{kremer2025practicalefficientquantumcircuit,dubal2025pauli} introduced Reinforcement Learning (RL) techniques to address these transpilation challenges, achieving near-optimal synthesis of various circuit types (Linear Function, Clifford, Permutation, Pauli Networks) while respecting device-native instruction sets and connectivity constraints. This approach demonstrated significant improvements in balancing circuit optimality and computational efficiency compared to traditional methods. 

Building upon this foundation, the current work focuses on extending the RL-based approach specifically for permutation circuits, which are fundamental components in many quantum algorithms including quantum Fourier transforms \cite{coppersmith2002approximate}, error correction codes \cite{fowler2012surface}, quantum simulation \cite{kivlichan2018quantum} and in applications such as the calculation of ground state energies or the preparation of time-evolved states in condensed matter systems \cite{babbush2018encoding,google2020hartree,robledo2024chemistry}. While our previous work demonstrated effective results for permutation circuit synthesis up to 65 qubits, that approach required training specialized models for each specific device topology.
In this paper, we present a generalist approach to RL-based quantum circuit transpilation for permutation circuits that overcomes this limitation. Rather than training separate models for different device topologies, our method employs a unified model architecture capable of adapting to various connectivity constraints. This advancement improves the practical utility of AI-assisted transpilation by eliminating the need to maintain multiple specialized models while maintaining performance comparable to the topology-specific models established in our previous work. Additionally, we provide comprehensive benchmarks comparing our against previous specialized RL models and Qiskit's TokenSwapper algorithm \cite{childs2019circuit,wagner2023improving}. The results demonstrate that the generic RL model presented here achieves similar performance to the specific models and maintains the advantage against heuristic methods, especially for circuit depth, but providing much more flexibility than the specialized models.

The paper is organized as follows: Section II presents the materials and methods used for this paper. Section III presents the main results related to training generic RL models for permutation circuits and how these result compare against using specialized models for each coupling map. Section IV discusses the results, presents some conclusions and future directions for this work.

\section{Materials and Methods}

\begin{figure*}[ht]
\includegraphics[width=1.0\textwidth]{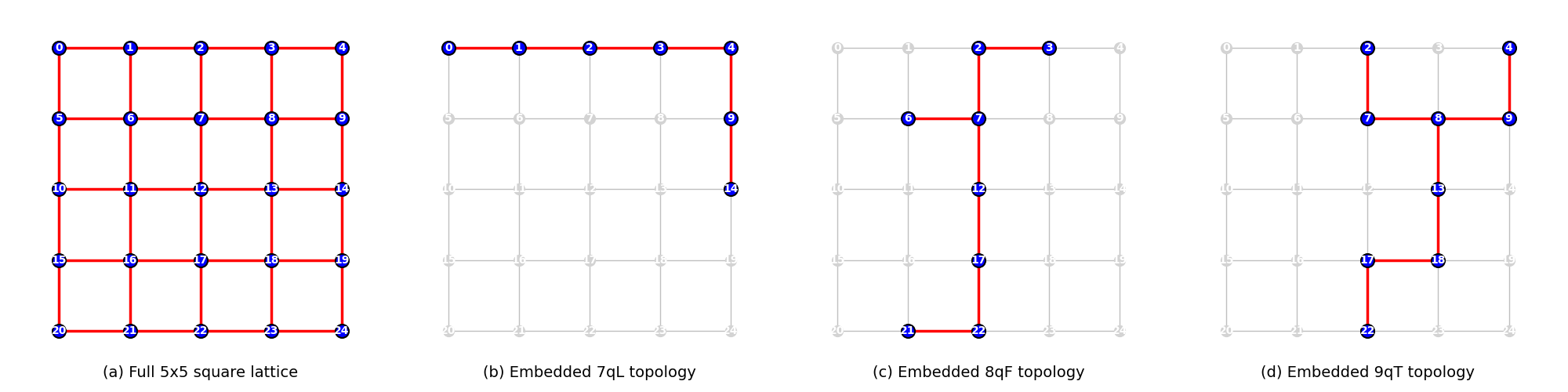}
\caption{Examples of topologies within the 5x5 square lattice, (a) Full 5x5 square lattice, (b) 7 qubits with "L" connectivity, (c) 8 qubits with "F" connectivity, (d) 9 qubits with "T" connectivity. }
\label{topologies_examples}
\end{figure*}

\begin{figure*}[ht]
\includegraphics[width=1.0\textwidth]{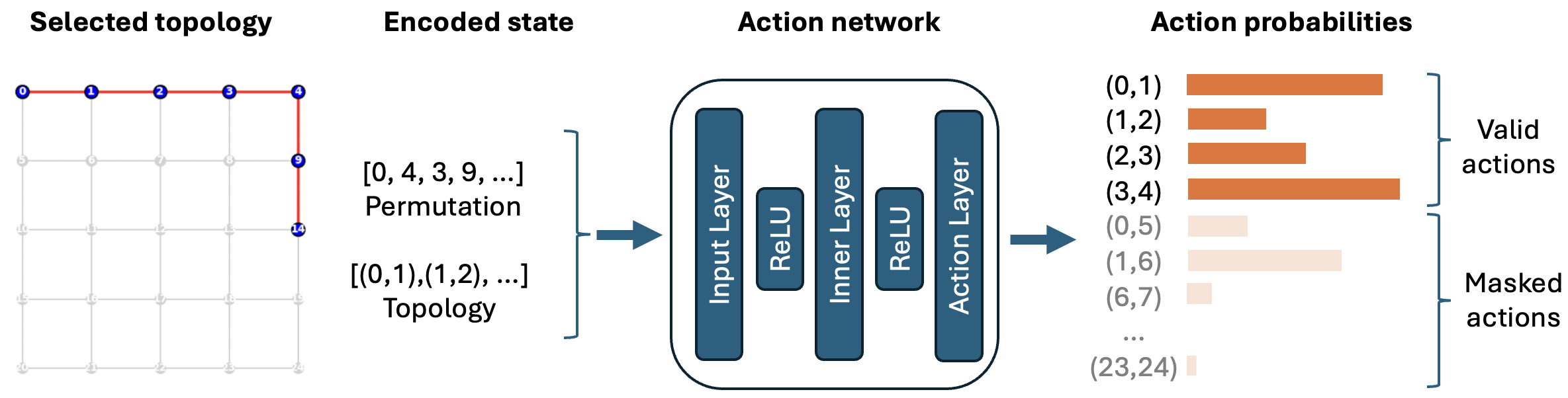}
\caption{Network architecture used as the RL policy. After a topology is selected, the permutation and the topology are encoded and passed to the network. We use a simple feedforward network with three layers. The network outputs the probability assigned to each possible action on the grid, but the sampling only happens among the actions that are allowed by the topology. }
\label{network_arch}
\end{figure*}

\subsection{Permutation Synthesis With Reinforcement Learning Across Generic Topologies}

Our approach applies RL to circuit synthesis by formulating the task as a sequential decision-making process. In this framework, synthesizing a permutation involves making a series of decisions: at each step, given an operator, the goal is to choose a swap from a predefined set of available operations, i.e. swaps between connected qubits. Applying this gate evolves the operator to a new state. 
Extending our work \cite{kremer2025practicalefficientquantumcircuit} for a more generalist approach, in this case, to train a model that works across different topologies, we use a general square lattice topology where we can set a specific topology. 
Starting with an initial operator (the target operation we wish to implement) and a topology that can be mapped onto a square lattice, the process continues for several steps until the identity operator is reached. The final circuit that implements the original permutation while following the selected topology is then obtained by inverting the sequence of gates. In Figure \ref{topologies_examples} we show the square lattice and a few topologies that we can embed in it.

At the heart of the method is an RL agent, which selects the appropriate gate at each step based on the current operator and the selected topology within the lattice. In order to prevent the model from applying gates that are not valid under the selected topology, we employ a method know as action masking \cite{ye2020mastering,vinyals2017starcraft}. In our network architecture the network technically outputs the action probabilities for all possible gates in the grid, but only the valid actions are considered when sampling from the distribution. This prevents the model from selecting swaps that are not allowed by the specified topology. The topology is also used as input to the network so that it allows specialized strategies to be learned. A depiction of the network architecture is shown in Figure \ref{network_arch}.

Like other AI-based techniques, our method includes a training phase during which the agent learns how to perform permutation synthesis. Once trained, the agent can synthesize circuits during an inference phase by applying the knowledge it has acquired.

\subsection{Training}

To train the RL agent, we follow a standard reinforcement learning pipeline in which the agent attempts to synthesize input operators and learns from the outcomes. As inputs we use the current operator and the selected topology. The latter is used as input and to mask the possible actions taken by the agent. In addition we enhance this process by progressively increasing the difficulty of the input operators as the agent improves, a strategy commonly referred to as curriculum learning in the RL literature \cite{narvekar2020curriculum}.

The agent receives feedback on its actions through a reward function. At each decision step, a reward is computed based on the immediate outcome of the chosen action. Our reward function consists of two main components:

\begin{enumerate}
    \item A large positive reward is given when the agent reaches the identity operator, signaling successful synthesis. This encourages the agent to take actions that complete the circuit.

    \item Small negative rewards (penalties) are applied for each gate used, promoting more efficient circuits with fewer gates and reduced depth. 
\end{enumerate}

During the training process we perform the following operations at each training step:

\begin{enumerate}
    \item Generate a batch of random input pairs, each one consisting of a target permutation and a specific topology. The topology is created starting from a random qubit and randomly selecting connections from there, in such a way that we end up with a connected sub-graph of the main lattice with a random number of qubits. Once we have the topology, we randomly create a permutation depending on the current difficulty level (see \cite{kremer2025practicalefficientquantumcircuit} for more details).
    
    \item Collect the needed observations and rewards at each step through the inference process for each (permutation target, topology) pair. The topology is used to mask invalid actions, to avoid getting actions not compatible with the connectivity restrictions in the inference process. 
    
    \item Use the collected data to update the network weights using the Proximal Policy Optimization (PPO) algorithm \cite{schulman2017proximal} as implemented in our own RL framework twisteRL\footnote{https://github.com/IBM/twisteRL}. Action masking is also used in this step.

    \item If the rate of input permutations successfully synthesized is above a specific threshold (typically 0.85) the difficulty is increased by 1.
\end{enumerate}

Once the RL agent is trained, we can synthesize a permutation for a specific topology by letting the RL agent take actions at each step. As we described, the RL agent consist of a neural network that takes the given permutation and topology as inputs (in a numerical representation) and outputs a log-likelihood for each of the possible actions, i.e. each possible swap allowed by the topology. It is important to note again that we use the input topology to mask the possible actions, making the probability to select a non-allowed action effectively zero.

In the inference process one can follow two main approaches: greedy and sampling. In the greedy approach we always select the action with the highest probability, whereas in the sampling one we sample from the probability distribution. At this point we see a fundamental difference with the topology specific models. While for the specific ones we can follow a greedy approach, for the generic model we need to use sampling to reach a high level of success rate when synthesizing permutations.

\section{Results}

\begin{figure*}[htp]
\centerline{\includegraphics[width=1.0\textwidth]{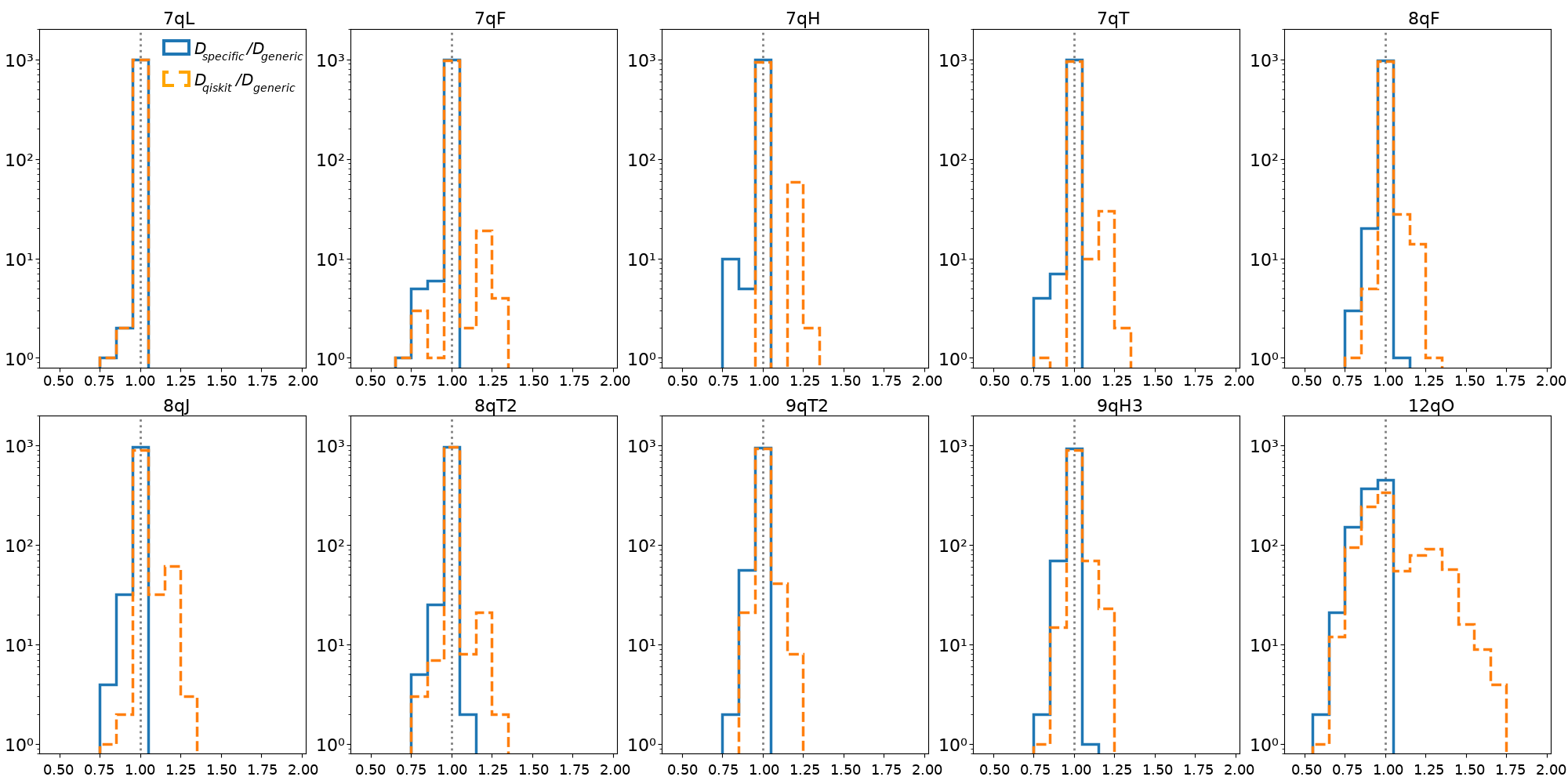}}
\caption{ Comparison of number of gates. Each plot presents a histogram with the ratios of gates for the different specific models. In blue, number of swaps obtained with the specific model over circuit depth obtained with the generic model. In orange, number of swaps obtained with Qiskit TokenSwaper algorithm over circuit depth obtained with the generic model. }
\label{comparison_with_models}
\end{figure*}

\begin{figure*}[htp]
\centerline{\includegraphics[width=1.0\textwidth]{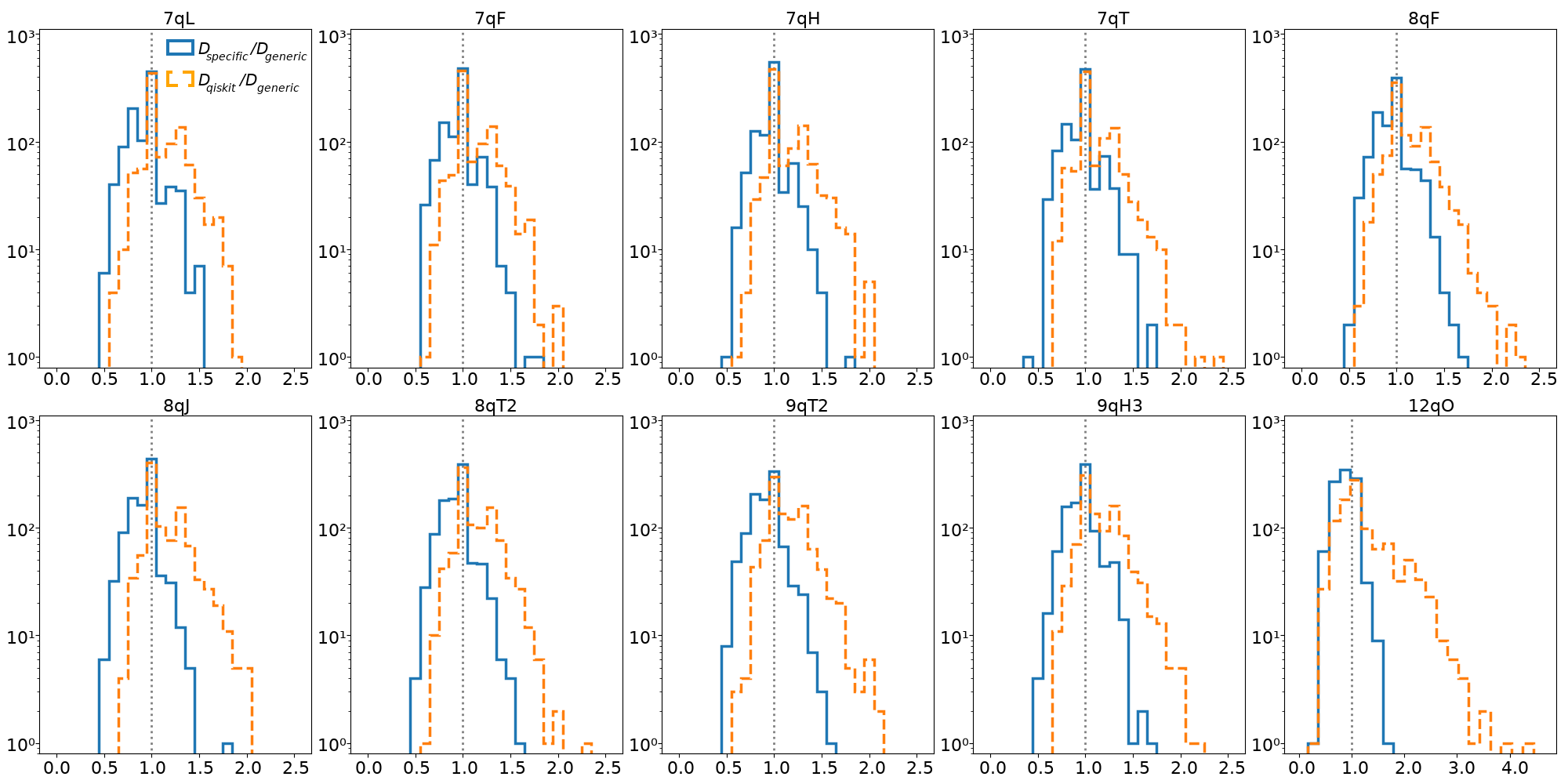}}
\caption{Comparison of circuit depths. Each plot presents a histogram with the ratios of depths for the different specific models. In blue, circuit depth obtained with the specific model over circuit depth obtained with the generic model. In orange, circuit depth obtained with Qiskit TokenSwaper algorithm over circuit depth obtained with the generic model. }
\label{depth_comparison_with_models}
\end{figure*}

In this section we analyze the performance of the generic topology models against models trained for specific topologies and Qiskit TokenSwapper algorithm.

We have chosen the following typical topologies 7qL, 7qF, 7qH, 7qT, 8qF, 8qJ, 8qT2, 9qT2, 9qH3 and 12qO, which can be embed in our generic 5x5 square lattice model. Find detailed information about these topologies in \cite{kremer2025practicalefficientquantumcircuit}'s supplementary information. Specialized models for these topologies have been previously trained following the approach we used in  \cite{kremer2025practicalefficientquantumcircuit}. To perform the comparison, first we create a set of random permutations for each of the specific topologies. We run both the generic and the specific model for each permutation to obtain the number of swaps implemented by each model to synthesize the permutation. In the case of the generic model, for each permutation, we randomly select a subgraph within the square lattice that matches the specific topology, and map the permutation to it. We run the comparisons running the inference following the greedy strategy for the specific models and the sampling one for the generic one, performing 10 runs.

\begin{table*}[t]
\begin{center}
\begin{tabular}{|c|c|c|c|c|c|c|c|c|c|c|}
\hline
&\multicolumn{5}{|c|}{\textbf{\textit{Specific model}}}& \multicolumn{5}{|c|}{\textbf{\textit{Qiskit TokenSwapper}}}\\
\hline
& \multicolumn{2}{|c|}{\textbf{\textit{$N/N^{generic}$}}} & \multicolumn{2}{|c|}{\textbf{\textit{$depth/depth^{generic}$}}} & \textbf{\textit{$t/t^{generic}$}} &\multicolumn{2}{|c|}{\textbf{\textit{$N/N^{generic}$}}} &\multicolumn{2}{|c|}{\textbf{\textit{$depth/depth^{generic}$}}} & \textbf{\textit{$t/t^{generic}$}}  \\
\hline
\textbf{\textit{Topology}} & $<0.95$ & $>1.05$ &  $<0.95$ & $>1.05$ & avg & $<0.95$ & $>1.05$ &  $<0.95$ & $>1.05$ & avg \\
\hline
7qL & 0.3\% & 0\% & 44\% & 11\% & 0.08 & 0.3\% & 0\% & 12\% & 44\% & 15\\
7qF & 1.2\% & 0\% & 36\% & 16\% & 0.08 & 0.5\% & 2.5\% & 10\% & 44\% & 10\\
7qH & 1.5\% & 0\% & 31\% & 14\% & 0.08 &0.0\% & 6.1\% & 8\% &  44\%& 10\\
7qT & 1.1\% & 0\% & 36\% & 17\% & 0.08& 0.1\% & 4.2\% & 12\% &  43\%& 10\\
\hline
8qF & 2.3\% & 0.1\% & 43\% & 17\% & 0.07 & 0.6\% & 4.3\% & 14\% &  50\%& 10\\
8qJ & 3.5\% & 0\% & 48\% & 8\% & 0.07  & 0.3\% & 9.6\% & 9\% &  50\%& 12\\
8qT2 & 3.0\% & 0.2\%& 49\% & 13\% & 0.07 & 1.0\% & 3.1\% & 11\% & 52\% & 11\\
\hline
9qT2 & 5.8\% & 0\% & 53\% & 13\% & 0.07 & 2.1\% & 4.9\% & 13\% & 58\% & 12\\
9qH3 & 7.2\% & 0.1\%& 41\% & 20\% & 0.07& 1.6\% & 9.2\% & 11\% & 58\% & 13\\
\hline
12qO & 54\% & 0\%& 67\% & 11\% & 0.05& 35\% & 31\% & 33\% & 48\% & 11\\
12qO FT & 17\% & 0.2\% & 48\% & 20\% & 0.07 & 10\% & 30\% & 17\% & 56\% & 13\\ 
\hline
\end{tabular}
\caption{Methods Comparison.}
\label{table_method_comp}
\end{center}
\end{table*}

In Figure \ref{comparison_with_models} we show the histograms for the ratios of number of swaps for a specific model over the number of swaps in the generic model $N_{specific}/N_{generic}$ and the number of gates obtained with Qiskit TokenSwapper (with 1000 trials) over the generic model. Values higher than one imply the generic model is performing better while the opposite is true for values lower than one. As observed in the figure, the generic model performs similarly to the specific models. Comparing to those specific models, the generic one achieves the same number of gates for a large fraction of input set ($> 90\%$) across all topologies, except for the topology 12qO, which we will discuss later. In the fraction of permutations synthesized where the generic model performs worse, the specific models typically achieves a 10\% - 20\% reduction in the number of gates. In a small fraction of cases we can see the generic model outperforms the specific model, generating  $\sim 10\%$ less gates. If we use the sampling strategy for both specific and generic models, the results are quite similar. When comparing to Qiskit TokenSwapper, results are significantly better, with the generic model outperforming the TokenSwapper in several topologies.
In Figure \ref{depth_comparison_with_models} we do an analog comparison but considering the depth of the circuits instead of the number of gates. Comparing with the specific models, we see that these perform slightly better on average, but in this case we also see that the generic model is better for a fraction of the input sample. In general, there is greater deviation from equality in both directions. Compared to Qiskit’s TokenSwapper, the generic model yields overall better results, although the values also exhibit more spread around the equality line in this case.

In terms of execution speed, the specific models are usually one order of magnitude faster than the generic model, while the Qiskit TokenSwapper (QiskitSDK v2.0) is one order of magnitudes slower than the generic model, while providing worse results on average.

In Table \ref{table_method_comp} we summarize the results comparing both specific models and Qiskit TokenSwapper to the generic model.

The case for the 12 qubits using a ring connectivity (O) is interesting. In this case the generic model clearly perform worse than the specific model, even though for half the test sample it synthesize circuits with the same number of gates. The probability of randomly selecting a ring topology in our training process is negligible. Thus, the model has never seen this topology in the training, and even though the model is able to synthesize the circuits, it clearly can improve. In this case we see that the specific model synthesizes circuits with less than $\sim 95\%$ gates than the generic model for $\sim55\%$ of the test sample. However, the generic model is capable of synthesizing circuits with at most $20\%$ more gates in approximately $\sim90\%$ of the cases, and at most $10\%$ more gates in $\sim70\%$. This means that, for a target permutation requiring 20 swaps, the generic model will produce a circuit with 24 swaps or fewer in around $90\%$ of instances, and with no more than 22 swaps in roughly $70\%$ of the cases.

\begin{figure}[t]
\centerline{\includegraphics[width=0.47\textwidth]{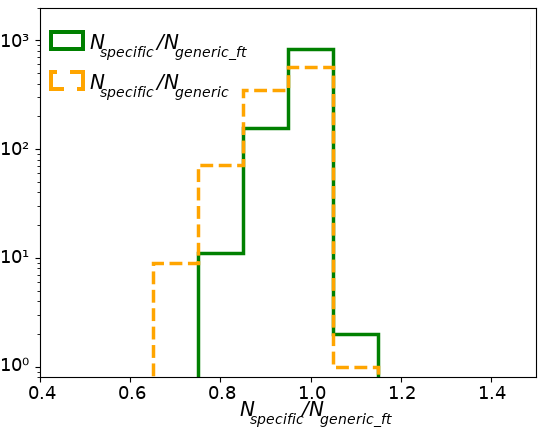}}
\caption{ Comparison of generic regular and fine tuned model for 12qO topology. In orange, number of gates for specific model over number of gates for generic model. In green, number of gates for specific model over number of gates for fine tuned generic model.}
\label{finetuned_12qO}
\end{figure}

To check our hypothesis that the problem lies in the lack of ring connectivity topologies in the training, we performed a simple fine tune of the generic model. We changed the algorithm that generates the random topologies and force the inclusion of a specific 12qO topology, and continue training the generic model from its current weights. In Figure \ref{finetuned_12qO} we show the comparison between the regular and fine tuned generic models. We can clearly see that if the generic model sees the 12qO connectivity during the training, the performance for that particular topology improves. In particular, the fraction of circuits for which the generic model generates more than 5\% more gates decreases from $\sim$54\% for the regular generic model to $\sim$17\% for the fine tuned generic model.

In Table \ref{table_model_parameters} we compare network parameters and size of the different models. The generic model is heavier in terms parameters and disk size compared to the specific models (2.5Mb versus $\sim$700Kb). However, the generic model alone is capable of perform similarly than the specific models in most cases and avoids the burden of having to train all the specific models, storing them and managing them, in case they are used as part of a any software library or service. 

\begin{table}[t]
\begin{center}
\begin{tabular}{|c|c|c|}
\hline
&\multicolumn{2}{|c|}{\textbf{\textit{Train configuration}}}\\
\hline
\textbf{\textit{\#Topology}}&\textbf{\textit{\#Network Parameters ($10^3$)}} & \textbf{\textit{Size (KB)}} \\ 
\hline
7qL & 159 & 623 \\
7qF & 159 & 623 \\
7qH & 159 & 623 \\
7qT & 159 & 623 \\
\hline
8qF & 167 & 654 \\
8qJ & 167 & 654 \\ 
8qT2 & 167 & 654 \\
\hline
9qT2 & 176 & 689 \\
9qH3 & 176 & 689 \\
\hline
12qO & 209 & 819 \\
\hline
Generic & 614 & 2500 \\
\hline
\end{tabular}
\caption{Model characteristics}
\label{table_model_parameters}
\end{center}
\end{table}

\section{Discussion}
In this work we have presented an RL-based method for the synthesis of permutations to generic topologies. The method addresses one of the main limitations of previous RL-based approaches \cite{kremer2025practicalefficientquantumcircuit, dubal2025pauli}, i.e. the need for training specific models for synthesis to different topologies. By including dynamic action masking during training, we show how one generic model can be trained to do synthesis of permutations to arbitrary topologies contained within a base coupling map, by just specifying the adequate mask during the synthesis and without the need to retrain.

In our benchmarks, we see that the generic RL model performs on pair with the specialized RL models for a wide range of topologies. We also observe that the model still works for topologies that are very scarce or non-existent in the training set, such as the 12 qubit ring, even if they initially perform worse than the specific models. We further show how these deficiencies can be improved by fine-tuning the generic model to improve on the specific topologies of interest.

Through this work we demonstrate that action masking can be an effective way to scale RL-based synthesis of permutations, where the network learns common patterns that help it perform effective synthesis across different topologies. A straightforward next step would be to validate if this approach can also be applied to synthesis of other types of circuits (e.g. Clifford, etc.). Another direction for future work would be to improve the network architecture in order to include generic transformations that allow to overcome the limitation of a fixed size grid, such as transformer-based architectures or graph neural networks.

\bibliography{references}

\begin{thebibliography}{25}
\providecommand{\natexlab}[1]{#1}

\bibitem[{Acampora and Schiattarella(2021)}]{acampora2021deep}
Acampora, G.; and Schiattarella, R. 2021.
\newblock Deep neural networks for quantum circuit mapping.
\newblock \emph{Neural Computing and Applications}, 33(20): 13723--13743.

\bibitem[{Babbush et~al.(2018)Babbush, Gidney, Berry, Wiebe, McClean, Paler,
  Fowler, and Neven}]{babbush2018encoding}
Babbush, R.; Gidney, C.; Berry, D.~W.; Wiebe, N.; McClean, J.; Paler, A.;
  Fowler, A.; and Neven, H. 2018.
\newblock Encoding electronic spectra in quantum circuits with linear T
  complexity.
\newblock \emph{Physical Review X}, 8(4): 041015.

\bibitem[{Botea, Kishimoto, and Marinescu(2018)}]{botea2018complexity}
Botea, A.; Kishimoto, A.; and Marinescu, R. 2018.
\newblock On the complexity of quantum circuit compilation.
\newblock In \emph{Proceedings of the International Symposium on Combinatorial
  Search}, volume~9, 138--142.

\bibitem[{Bravyi, Latone, and Maslov(2022)}]{bravyi20226}
Bravyi, S.; Latone, J.~A.; and Maslov, D. 2022.
\newblock 6-qubit optimal Clifford circuits.
\newblock \emph{npj Quantum Information}, 8(1): 79.

\bibitem[{Childs, Schoute, and Unsal(2019)}]{childs2019circuit}
Childs, A.~M.; Schoute, E.; and Unsal, C.~M. 2019.
\newblock Circuit transformations for quantum architectures.
\newblock \emph{arXiv preprint arXiv:1902.09102}.

\bibitem[{Coppersmith(2002)}]{coppersmith2002approximate}
Coppersmith, D. 2002.
\newblock An approximate Fourier transform useful in quantum factoring.
\newblock \emph{arXiv preprint quant-ph/0201067}.

\bibitem[{Dubal et~al.(2025)Dubal, Kremer, Martiel, Villar, Wang, and
  Cruz-Benito}]{dubal2025pauli}
Dubal, A.; Kremer, D.; Martiel, S.; Villar, V.; Wang, D.; and Cruz-Benito, J.
  2025.
\newblock Pauli Network Circuit Synthesis with Reinforcement Learning.
\newblock \emph{arXiv preprint arXiv:2503.14448}.

\bibitem[{F{\"o}sel et~al.(2021)F{\"o}sel, Niu, Marquardt, and
  Li}]{fosel2021quantum}
F{\"o}sel, T.; Niu, M.~Y.; Marquardt, F.; and Li, L. 2021.
\newblock Quantum circuit optimization with deep reinforcement learning.
\newblock \emph{arXiv preprint arXiv:2103.07585}.

\bibitem[{Fowler et~al.(2012)Fowler, Mariantoni, Martinis, and
  Cleland}]{fowler2012surface}
Fowler, A.~G.; Mariantoni, M.; Martinis, J.~M.; and Cleland, A.~N. 2012.
\newblock Surface codes: Towards practical large-scale quantum computation.
\newblock \emph{Physical Review A—Atomic, Molecular, and Optical Physics},
  86(3): 032324.

\bibitem[{Kivlichan et~al.(2018)Kivlichan, McClean, Wiebe, Gidney,
  Aspuru-Guzik, Chan, and Babbush}]{kivlichan2018quantum}
Kivlichan, I.~D.; McClean, J.; Wiebe, N.; Gidney, C.; Aspuru-Guzik, A.; Chan,
  G. K.-L.; and Babbush, R. 2018.
\newblock Quantum simulation of electronic structure with linear depth and
  connectivity.
\newblock \emph{Physical review letters}, 120(11): 110501.

\bibitem[{Kremer et~al.(2025)Kremer, Villar, Paik, Duran, Faro, and
  Cruz-Benito}]{kremer2025practicalefficientquantumcircuit}
Kremer, D.; Villar, V.; Paik, H.; Duran, I.; Faro, I.; and Cruz-Benito, J.
  2025.
\newblock Practical and efficient quantum circuit synthesis and transpiling
  with Reinforcement Learning.
\newblock \emph{arXiv preprint arXiv:2405.13196}.

\bibitem[{Li, Ding, and Xie(2019)}]{li2019tackling}
Li, G.; Ding, Y.; and Xie, Y. 2019.
\newblock Tackling the qubit mapping problem for NISQ-era quantum devices.
\newblock In \emph{Proceedings of the twenty-fourth international conference on
  architectural support for programming languages and operating systems},
  1001--1014.

\bibitem[{Murali et~al.(2019)Murali, Baker, Javadi-Abhari, Chong, and
  Martonosi}]{murali2019noise}
Murali, P.; Baker, J.~M.; Javadi-Abhari, A.; Chong, F.~T.; and Martonosi, M.
  2019.
\newblock Noise-adaptive compiler mappings for noisy intermediate-scale quantum
  computers.
\newblock In \emph{Proceedings of the twenty-fourth international conference on
  architectural support for programming languages and operating systems},
  1015--1029.

\bibitem[{Narvekar et~al.(2020)Narvekar, Peng, Leonetti, Sinapov, Taylor, and
  Stone}]{narvekar2020curriculum}
Narvekar, S.; Peng, B.; Leonetti, M.; Sinapov, J.; Taylor, M.~E.; and Stone, P.
  2020.
\newblock Curriculum learning for reinforcement learning domains: A framework
  and survey.
\newblock \emph{Journal of Machine Learning Research}, 21(181): 1--50.

\bibitem[{Nielsen and Chuang(2010)}]{nielsen2010quantum}
Nielsen, M.~A.; and Chuang, I.~L. 2010.
\newblock \emph{Quantum computation and quantum information}.
\newblock Cambridge university press.

\bibitem[{Niu et~al.(2019)Niu, Boixo, Smelyanskiy, and
  Neven}]{niu2019universal}
Niu, M.~Y.; Boixo, S.; Smelyanskiy, V.~N.; and Neven, H. 2019.
\newblock Universal quantum control through deep reinforcement learning.
\newblock \emph{npj Quantum Information}, 5(1): 33.

\bibitem[{Olle, Yevtushenko, and Marquardt(2025)}]{olle2025scaling}
Olle, J.; Yevtushenko, O.~M.; and Marquardt, F. 2025.
\newblock Scaling the Automated Discovery of Quantum Circuits via Reinforcement
  Learning with Gadgets.
\newblock \emph{arXiv preprint arXiv:2503.11638}.

\bibitem[{Preskill(2018)}]{preskill2018quantum}
Preskill, J. 2018.
\newblock Quantum computing in the NISQ era and beyond.
\newblock \emph{Quantum}, 2: 79.

\bibitem[{Quantum et~al.(2020)Quantum, Collaborators*†, Arute, Arya, Babbush,
  Bacon, Bardin, Barends, Boixo, Broughton, Buckley et~al.}]{google2020hartree}
Quantum, G.~A.; Collaborators*†; Arute, F.; Arya, K.; Babbush, R.; Bacon, D.;
  Bardin, J.~C.; Barends, R.; Boixo, S.; Broughton, M.; Buckley, B.~B.; et~al.
  2020.
\newblock Hartree-Fock on a superconducting qubit quantum computer.
\newblock \emph{Science}, 369(6507): 1084--1089.

\bibitem[{Robledo-Moreno et~al.(2024)Robledo-Moreno, Motta, Haas,
  Javadi-Abhari, Jurcevic, Kirby, Martiel, Sharma, Sharma, Shirakawa
  et~al.}]{robledo2024chemistry}
Robledo-Moreno, J.; Motta, M.; Haas, H.; Javadi-Abhari, A.; Jurcevic, P.;
  Kirby, W.; Martiel, S.; Sharma, K.; Sharma, S.; Shirakawa, T.; et~al. 2024.
\newblock Chemistry beyond exact solutions on a quantum-centric supercomputer.
\newblock \emph{arXiv preprint arXiv:2405.05068}.

\bibitem[{Schulman et~al.(2017)Schulman, Wolski, Dhariwal, Radford, and
  Klimov}]{schulman2017proximal}
Schulman, J.; Wolski, F.; Dhariwal, P.; Radford, A.; and Klimov, O. 2017.
\newblock Proximal policy optimization algorithms.
\newblock \emph{arXiv preprint arXiv:1707.06347}.

\bibitem[{Tannu and Qureshi(2019)}]{tannu2019not}
Tannu, S.~S.; and Qureshi, M.~K. 2019.
\newblock Not all qubits are created equal: A case for variability-aware
  policies for NISQ-era quantum computers.
\newblock In \emph{Proceedings of the twenty-fourth international conference on
  architectural support for programming languages and operating systems},
  987--999.

\bibitem[{Vinyals et~al.(2017)Vinyals, Ewalds, Bartunov, Georgiev, Vezhnevets,
  Yeo, Makhzani, K{\"u}ttler, Agapiou, Schrittwieser
  et~al.}]{vinyals2017starcraft}
Vinyals, O.; Ewalds, T.; Bartunov, S.; Georgiev, P.; Vezhnevets, A.~S.; Yeo,
  M.; Makhzani, A.; K{\"u}ttler, H.; Agapiou, J.; Schrittwieser, J.; et~al.
  2017.
\newblock Starcraft ii: A new challenge for reinforcement learning.
\newblock \emph{arXiv preprint arXiv:1708.04782}.

\bibitem[{Wagner et~al.(2023)Wagner, B{\"a}rmann, Liers, and
  Weissenb{\"a}ck}]{wagner2023improving}
Wagner, F.; B{\"a}rmann, A.; Liers, F.; and Weissenb{\"a}ck, M. 2023.
\newblock Improving quantum computation by optimized qubit routing.
\newblock \emph{Journal of Optimization Theory and Applications}, 197(3):
  1161--1194.

\bibitem[{Ye et~al.(2020)Ye, Liu, Sun, Shi, Zhao, Wu, Yu, Yang, Wu, Guo
  et~al.}]{ye2020mastering}
Ye, D.; Liu, Z.; Sun, M.; Shi, B.; Zhao, P.; Wu, H.; Yu, H.; Yang, S.; Wu, X.;
  Guo, Q.; et~al. 2020.
\newblock Mastering complex control in moba games with deep reinforcement
  learning.
\newblock In \emph{Proceedings of the AAAI conference on artificial
  intelligence}, volume~34, 6672--6679.

\end{thebibliography}

\end{document}